\documentclass[conference]{IEEEtran}
\usepackage{alltt                                    
          , multirow
          , booktabs
          , listings
          , graphicx
          ,float
	,cite
          ,verbatim
         ,mathtools
	,url
	,amsmath
}
\usepackage[table]{xcolor}
\usepackage[numbers]{natbib}     
\usepackage{syntax}
\usepackage{algorithmic, algorithm}
\usepackage{enumitem}
\usepackage{threeparttable}
\usepackage[T1]{fontenc}

\usepackage{expl3}
\ExplSyntaxOn
\newcommand\latinabbrev[1]{
  \peek_meaning:NTF . {
    #1\@}%
  { \peek_catcode:NTF a {
      #1., \@ }%
    {#1., \@}}}
\ExplSyntaxOff

\newcommand{\CASE}[1]{\STATE \textbf{case} #1\textbf{:} \begin{ALC@g}}
\newcommand{\ENDCASE}{\end{ALC@g}}

\newcommand{\DEFAULT}{\STATE \textbf{default:} \begin{ALC@g}}
\newcommand{\ENDDEFAULT}{\end{ALC@g}}
\newcommand{\DEFAULTLINE}[1]{\STATE \textbf{default:} }
\let\footnotesize\scriptsize

\newsavebox{\supbox}
\newcommand{\bsup}{\begin{lrbox}{\supbox}$\tt\scriptstyle}
\newcommand{\esup}{$\end{lrbox}{}^{\usebox{\supbox}}}
\def\eg{\latinabbrev{e.g}}
\def\ie{\latinabbrev{i.e}}

\definecolor{lightpurple}{rgb}{0.8,0.8,1}
\definecolor{codebg}{RGB}{255,255,255}
\definecolor{commentcolor}{RGB}{11,140,11}
\definecolor {red}{rgb}{255,0,0}
\newcommand{\hl}{\makebox[0pt][l]{\color{lightgray}\rule[-2.5pt]{0.80\linewidth}{9pt}}}

\begin{document}
%

\title{SurfClipse: Context-Aware Meta Search in the IDE \vspace{-0.3cm}}

%
%
%

\author{\IEEEauthorblockN{Mohammad Masudur Rahman~~~~~Chanchal K. Roy }
\IEEEauthorblockA{Department of Computer Science, University of Saskatchewan, Canada \\
\{masud.rahman, chanchal.roy\}@usask.ca}
}



%
\maketitle
\begin{abstract}
Despite various debugging supports of the existing IDEs for programming errors and exceptions, software developers often look at web for working solutions or any up-to-date information.
Traditional web search does not consider the context of the problems that they search solutions for, and thus it often does not help much in problem solving.
In this paper, we propose a context-aware meta search tool, \emph{SurfClipse}, that analyzes an encountered exception and its context in the IDE, and recommends not only suitable search queries but also relevant web pages for the exception (and its context).
The tool collects results from three popular search engines and a programming Q \& A site against the exception in the IDE, refines the results for relevance against the context of the exception, and then ranks them before recommendation. 
It provides two working modes--\emph{interactive} and \emph{proactive} to meet the versatile needs of the developers, and one can browse the result pages using a customized embedded browser provided by the tool.

Tool page: www.usask.ca/$\sim$masud.rahman/surfclipse
\end{abstract}




\IEEEpeerreviewmaketitle


\section{Introduction}
Although existing IDEs (\eg\ Eclipse, Net Beans, Visual Studio) are equipped with various debugging supports for programming errrors and exceptions, software developers often look into the web for working solutions and for any up-to-date information.
According to the study of \citet{twostudy}, developers spend about 19\% of their development time in web surfing.
Traditional web search does not consider the \emph{context} (\ie\ surroundings, circumferences) of the programming problems, 
and involves the developers in \emph{trial and error} based query selection and search, which are time-consuming and counter-productive.
However, tool support through \emph{relevant query suggestion} and \emph{context-aware ranking} of search results can greatly benefit them in this regard, and this paper focuses on these two research problems.

There exist several studies \cite{context, seahawk} that attempt to address similar research problems.
\citet{context} propose an IDE-based recommendation system that recommends relevant StackOverflow posts for programming errors and exceptions. They extract a number of question and answer posts from StackOverflow data dump, and 
suggest those question posts that contain stack traces similar to that of an encountered exception in the IDE.
\citet{seahawk} propose \emph{Seahawk}, an Eclipse plugin, that analyzes the \emph{context code} (\ie\ code under development) of the current programming task and recommends relevant StackOverflow posts in the IDE.
Although these approaches have their inherent strengths, they also suffer from several limitations.
First, they consider only one source-- StackOverflow Q \& A site, for information, and thus the search scope is limited.
Second, the developed corpus is static in nature and also subjected to the availability of the data dump. 
Third, they only consider either stack trace or source code under development as the \emph{context} of a programming problem, which is partial (\ie\ incomplete) and often does not help much.
For example, the approach by \citeauthor{context} does not consider the code segment that triggers an exception and thus recommends solutions which might be non-applicable or even irrelevant to the code of interest given that the same exception could be triggered from different code context. Similarly, \emph{Seahawk} \cite{seahawk} cannot recommend properly for the programming tasks associated with errors and exceptions as it does not analyze the stack traces reported by the IDE.


\begin{figure*}[!t]
\centering
\includegraphics[width=7.1in ]{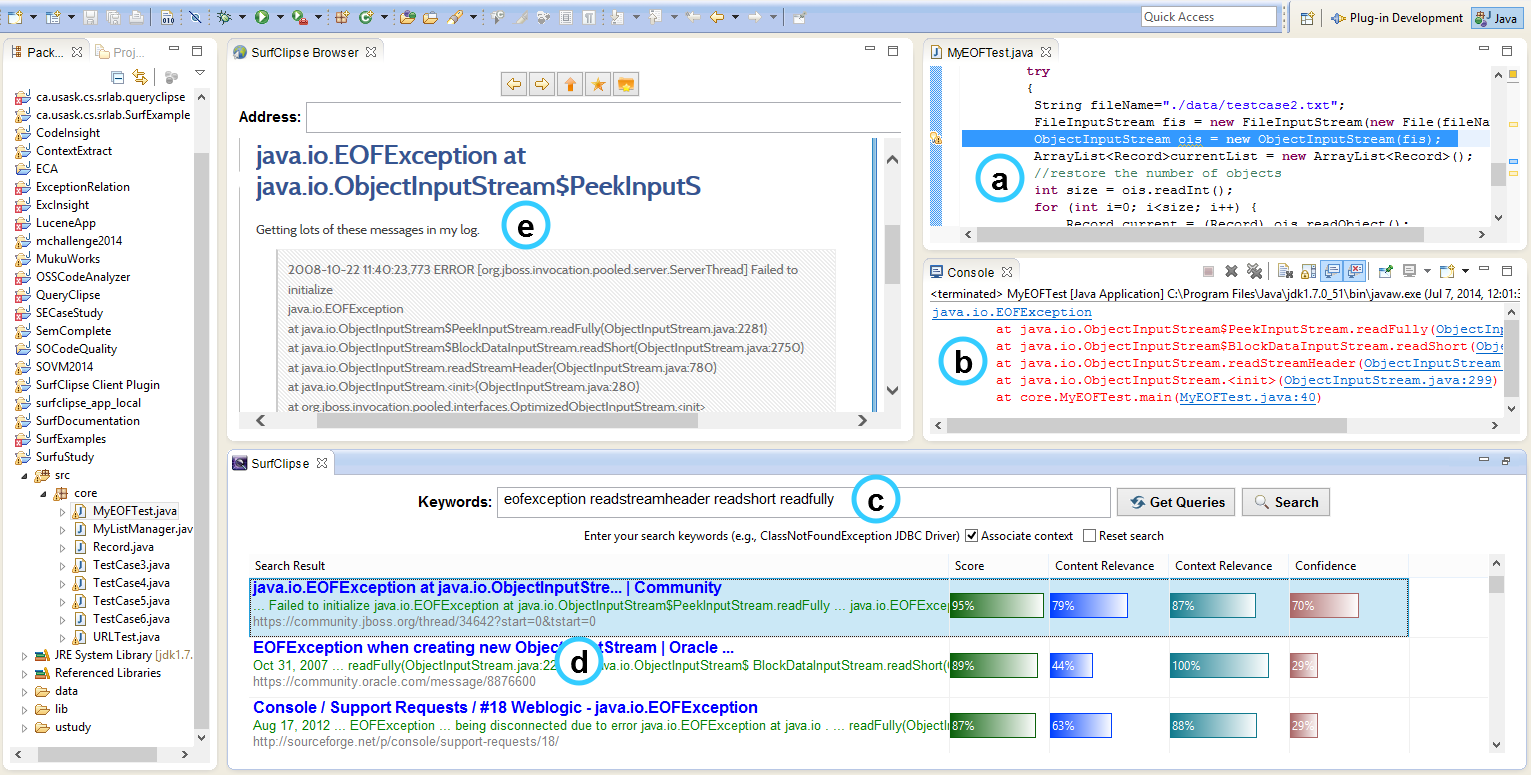}
\vspace{-.4cm}
\caption{SurfClipse User Interface}
\label{fig:ide}
\vspace{-.5cm}
\end{figure*}

In this paper, we propose a \emph{context-aware} \emph{meta search} tool, \emph{SurfClipse}, for the encountered programming errors and exceptions in the IDE that 
not only provides a complete web search solution but also addresses the concerns identified with the existing approaches \cite{context, seahawk}.
We package the solution as an Eclipse plugin \cite{scp}, which collects search results from a remotely hosted web service \cite{scp} and displays them within the IDE. 
Technically, the tool works both as a search query recommender and a meta search engine.
Once a developer encounters an exception in the IDE, the tool captures and analyzes the technical details (\ie\ stack trace and context code) of the exception, and recommends a list of relevant search queries. 
The developer selects a query from the list, and the tool collects results from three reliable search engines-- Google, Bing and Yahoo and a popular programming Q \& A site, StackOverflow, against the query.
It then analyzes, refines and ranks the results against not only the exception but also its context in the IDE before recommendation.
To summarize, our tool provides the following features to support developers in problem solving:
\begin{enumerate}[topsep=.2ex]
\item the proposed tool captures technical details of an encountered exception, and recommends a ranked list of suitable search queries that can be used both with \emph{SurfClipse} and traditional web search,
\item ranks the result pages adopting a \emph{context-aware} approach so that the pages are relevant not only to the encountered exception but also to its context in the IDE,
\item provides two working modes-- \emph{interactive} and \emph{proactive}, to meet the versatile needs of different developers and different task scenarios.
\item to ensure a broader search space, exploits the search and ranking algorithms of three popular search engines and a programming Q \& A site through their API endpoints,
\item exploits a dynamic source (\eg\ API endpoints) compared to static source (\eg\ data dump) by existing approaches \cite{context,seahawk} for StackOverflow data, which makes the most recent and relevant StackOverflow posts available for recommendation.
\end{enumerate}

\noindent
While this paper focuses on the tool aspect of our approach, we refer the readers to the original paper \cite{surf} for further details.


\section{SurfClipse}\label{sec:surf}
Fig. \ref{fig:ide} shows the user interface of \emph{SurfClipse}, where we contribute in (c) search panel, (d) result panel and (e) browser panel of the interface.
This section discusses different technical features provided by our tool.

\textbf{(1) Working Modes}: \emph{SurfClipse} works in two modes--\emph{interactive} and \emph{proactive}. In case of \emph{interactive} mode, a user (\eg\ a developer) generally initiates the search by selecting an exception from \emph{Console View} in the IDE or a search query (representing the exception and its context) from the recommendation list, whereas the tool itself initiates the search process in case of \emph{proactive} mode. 
Once the tool is properly installed, it provides several main menu and context-menu based command options, which can be used to initiate the tool environment or to change the working modes.

\textbf{(2) Automated Supports with Search Queries}: 
Both the context code (\eg\ Fig. \ref{fig:ide}-(a)) that triggers an exception, and the stack trace (\eg\ Fig. \ref{fig:ide}-(b)) reported by the IDE contain overwhelming information, and 
developers often face difficulties in choosing a suitable search query from such information.
\emph{SurfClipse} provides automated supports in this regard, and helps them choosing queries from two options--\emph{recommendation list} and \emph{stack trace graph}.

\begin{figure}
\centering
\includegraphics[width=3.4in ]{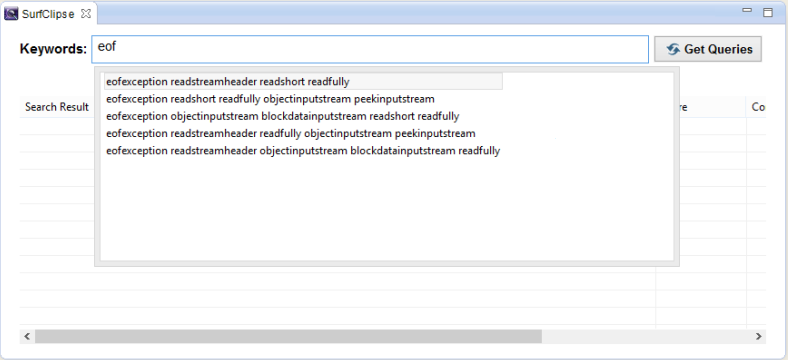}
\vspace{-.2cm}
\caption{Search Query Recommendation (Interactive Mode)}
\label{fig:qrecomm}
\vspace{-.5cm}
\end{figure}

\textbf{Query Recommendation}: In this case, the tool analyzes both stack trace and context code of the exception, and recommends a ranked list of five suitable search queries for the exception (Fig. \ref{fig:qrecomm}).

\begin{figure}
\centering
\includegraphics[width=2.2in ]{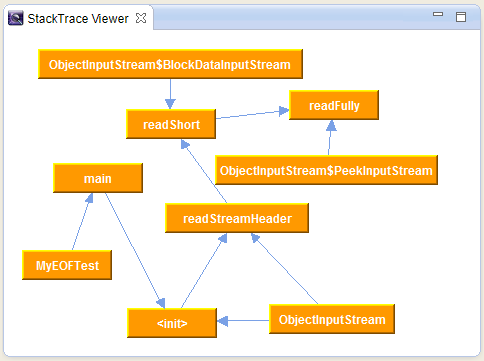}
\vspace{-.2cm}
\caption{Stack Trace Token Graph}
\label{fig:sgraph}
\vspace{-.2cm}
\end{figure}

\textbf{Stack Trace Token Graph}: In this case, the tool extracts important tokens (\eg\ class name, method name) from the stack trace (\eg\ Fig. \ref{fig:ide}-(b)), and develops a \emph{token graph} (\eg\ Fig. \ref{fig:sgraph}).
In the graph, tokens are represented as nodes, and the implied relationships (\eg\ class to method static relations, method call sequences) among the tokens are represented as connecting edges.
Thus the graph visualizes the relative importance of different tokens in terms of their connectivity, and the developers can get important hints about the useful query tokens in the trace information.

\textbf{(3) Context-Aware Web Search}: \emph{SurfClipse} provides three options to conduct web search within the IDE--\emph{proactive search}, \emph{context-menu based search (\ie\ interactive mode)} and \emph{keyword search (\ie\ interactive mode)}. 

\textbf{Proactive Search}: When \emph{SurfClipse} is set to \emph{proactive mode}, it automatically detects an encountered exception in the IDE. In this mode, the tool constantly monitors the \emph{Console View} for a stack trace using regular expressions.
Upon detection, it collects other details-- an auto-generated query and the context code of the exception, and initiates the search.

\begin{figure}[!t]
\centering
\includegraphics[width=2.2in ]{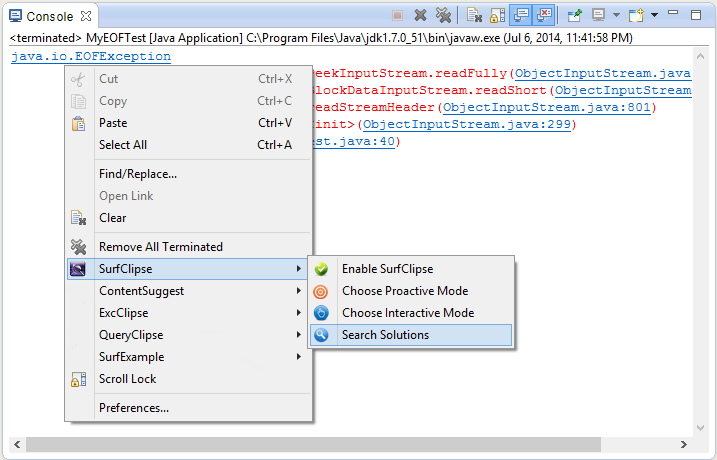}
\vspace{-.3cm}
\caption{Context-menu Based Search (Interactive Mode)}
\label{fig:cmsearch}
\vspace{-.5cm}
\end{figure}

\textbf{Context-Menu Based Search}: The tool provides a context-menu based search option, and a developer can literally select any phrase in the IDE (from \emph{Editor View} and \emph{Console View}), and perform web search. 
More importantly, she can select the exception in the \emph{Console View}, and conduct the search (\eg\ Fig. \ref{fig:cmsearch}). Once initiated, the tool captures necessary details from the IDE, performs the search, and collects the results. 

\textbf{Keyword Search}: Given that a user might be interested in refining the auto-generated search query or in a more traditional way of search, the tool provides a keyword-based search feature (\eg\ Fig. \ref{fig:ide}-(c)). The search is complemented with search query suggestion through auto-completion. The user can also configure whether the search should be a \emph{keyword matching only} (\ie\ does not refine the results against the \emph{exception context})  or a \emph{context-aware} one through \emph{Associate context} option (\eg\ Fig. \ref{fig:ide}-(c)).

\textbf{(4) Search Results \& Browsing}: Once a search request is made for an exception, the tool collects results in a non-intrusive way (\ie\ without freezing the IDE), and displays them within the IDE (\eg\ Fig. \ref{fig:ide}-(d)).
It also shows the metric details--\emph{content relevance, context relevance} and \emph{search engine confidence} of each result page through visualization, which helps one to choose the right (most relevant) page for browsing.
One can select a page from the result panel, and browse it easily using a customized browser widget (\eg\ Fig. \ref{fig:ide}-(e)) provided by the tool.

\section{A Use Case Scenario}\label{sec:usecase}
By means of a use case scenario, we attempt to explain how \emph{SurfClipse} can help a software developer in solving problems related to programming errors and exceptions within the IDE.

\begin{lstlisting}[label=lst:code, escapechar=@, frame=bt, belowskip=0em, numbers=left, aboveskip=1em,  xleftmargin=2.2em,firstnumber=20, xrightmargin=1.2em, float=tb,  caption={Working Code Example}]
 List<String> myList = new ArrayList<String>();
		String[] items={"apple","orange","banana",
                       "mango","grape"};
		for(String item:items){
			myList.add(item);  }
        //deleting a particular item from the list
        Iterator<String> it = myList.iterator();
        while(it.hasNext()){
           @\hl{String value = it.next();}@
            if(value.equals("banana")) 
             myList.remove(value);  }
\end{lstlisting}

Suppose a developer, Alice, is performing unit testing on a piece of code which searches for a specific item in the list and deletes the item when found (Listing \ref{lst:code}). During testing, she encounters a \emph{ConcurrentModification} exception, and gets a stack trace (Listing \ref{lst:exception}) reported by the IDE. She is not pretty familiar with that exception; however, from the stack trace, she identifies the source line (\ie\ Line 28, highlighted) triggering the exception, and she also assumes that the problem is something related to \emph{ArrayList}.

\begin{lstlisting}[label=lst:exception, frame=bt, aboveskip=0em, belowskip=-2em, float=t, numbers=left, xleftmargin=2.2em, xrightmargin=1.2em,  caption={Stack trace of ConcurrentModificationException}]
   Exception in thread "main" java.util.ConcurrentModificationException
   at java.util.ArrayList$Itr.checkForComodification(Unknown Source)
   at java.util.ArrayList$Itr.next(Unknown Source)
   at core.MyListManager.main(MyListManager.java:28)
\end{lstlisting}

She attempts to solve the issue, and at some point, she decides to perform web search for a solution or more helpful information. Now she faces several challenges--(1) How to develop an effective and appropriate search query for the current exception? (2) How to analyze and include the \emph{context} of the encountered exception during search? and finally (3) How to choose a working solution for the current exception from the search results? The traditional web search does not help her to overcome those challenges;
or if it does somehow, those are often not sufficient enough for problem solving.

Now let us assume that Alice has installed \emph{SurfClipse} in her IDE, and she encounters the same exception during testing. Our tool provides her with several options such as \emph{proactive search, context-menu based search} and \emph{keyword search}.
Suppose, she is interested in the third option-- keyword search. During this search, she can easily choose a search query from the recommended list, and the tool returns the top 30 result pages with the rationale (\ie\ metric details) behind their selection in the ranked list. 
Thus, in order to overcome the challenges that Alice faces with traditional search, our tool 
(1) helps her to develop a search query by either automatic suggestion (\eg\ Fig. \ref{fig:qrecomm}) or stack trace graph visualization (\eg\ Fig. \ref{fig:sgraph}),
(2) automatically captures the exception and its context (\ie\ stack trace and context code) from the IDE, and associates them with the search,
(3) returns results which are relevant not only to the exception but also to its context, and explains the rationale behind the selection of each result through metric details visualization (\eg\ Fig. \ref{fig:ide}-(d)).
Furthermore, the results are collected from four reliable and popular sources. 
\begin{figure}[!t]
\centering
\includegraphics[width=3.45in ]{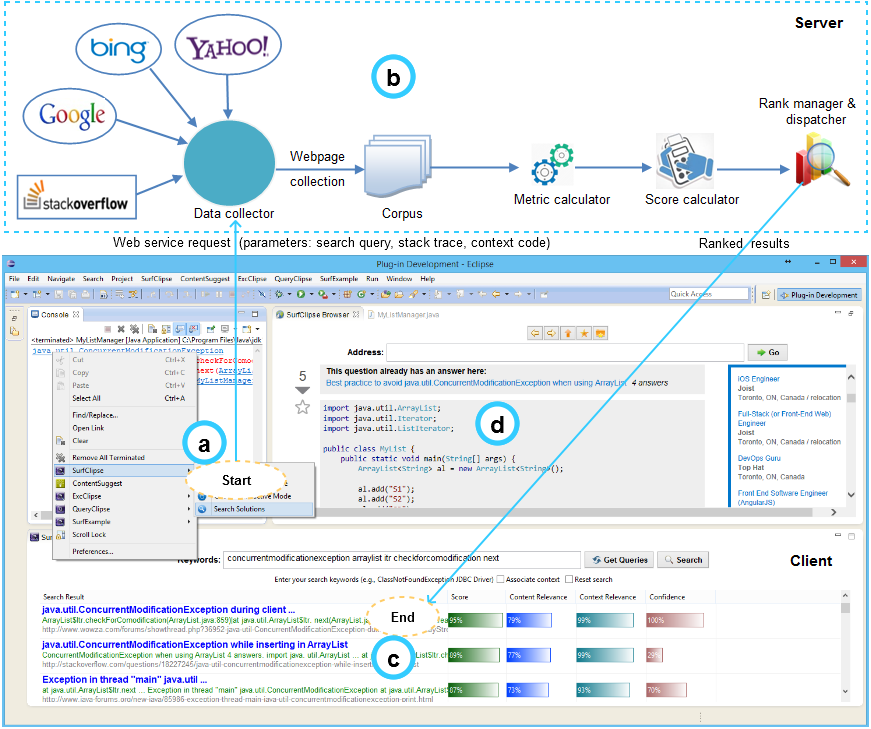}
\caption{Schematic Diagram of \emph{SurfClipse}}
\label{fig:sysdiag}
\vspace{-.2cm}
\end{figure}

\section{Working Methodology}\label{sec:methodology}
Fig. \ref{fig:sysdiag} shows the schematic diagram of the proposed tool.
This section discusses the internal structures and working methodologies of the tool in brief, while we refer the readers elsewhere \cite{surf} for details. 

\textbf{Search Query Formulation}: Once an exception occurs, we analyze both stack trace and context code of the exception in order to extract suitable query tokens. We develop a token graph (\eg\ Fig. \ref{fig:sgraph}) from the stack trace, where the connectivity of a token is based on its implied relationships (\eg\ class to method static relations) with other tokens. We consider the connectivity as a measure of token's importance, and exploit a graph-based term weighting algorithm (\ie\ a variation of PageRank algorithm) in order to determine the \emph{weight} of each token. We estimate the \emph{degree of interest} \cite{context} of each token, and also calculate the \emph{frequency} of the token in the context code. We then normalize each of the three metrics, and calculate the final score (\ie\ importance) of each token in the graph. Finally, we choose the top scored five tokens, and combine each three of them to formulate a list of search queries. Each of the queries essentially gets a score based on its token scores, and then the queries are also ranked for recommendation.

\textbf{Data Collection \& Context-Aware Ranking}:
The proposed tool follows a client-server architecture and it has two major entities-- Eclipse plugin (client) (Fig. \ref{fig:sysdiag}-(a, c, d)) and web service provider (server) (Fig. \ref{fig:sysdiag}-(b)). 
Once a developer selects an encountered exception from \emph{Console View} in the IDE, the client plugin collects associated context-- \emph{stack trace} and \emph{context code},
and generates a web search request to the service provider \cite{scp} (\eg\ Fig. \ref{fig:sysdiag}-(b)). The provider module
collects results from three search engines-- \emph{Google}, \emph{Bing} and \emph{Yahoo} and \emph{StackOverflow} Q \& A site against the client provided search query, and develops a dynamic corpus containing 100-120 result pages.
It then analyzes the content of the pages, and checks if they discuss the exception of interest and the discussed exception belongs to a programming context similar to the one in the IDE and so on. 
We use title and textual content of the page to determine its content level relevance, whereas we exploit the content of \emph{<code>,<pre>} and \emph{<blockquote>} tags in the page to determine its context relevance. Those tags generally contain the \emph{context} (\ie\ stack traces and code segments) associated with the discussed exception in the page. Finally, each of the result pages is ranked based on its \emph{content relevance}, \emph{context relevance}, \emph{search engine confidence} and \emph{popularity} (metric details can be found elsewhere \cite{surf}). The service provider module then returns the top 30 results, and the client plugin displays them in the IDE with corresponding metric details (Fig. \ref{fig:sysdiag}-(c)).

\section{Performance}\label{sec:performance}
In order to evaluate the recommended queries by \emph{SurfClipse}, we conducted a user study with five participants (graduate students) using five problem solving scenarios. Each of the participants solves the five exceptions, and we collect the queries they use for web search for each of the exceptions. We then compare  the recommended queries by our tool with those queries for the same exception using \emph{pyramid score} \cite{pyramid}. 
The metric determines if an auto-generated query resembles with a set of manually prepared search queries for the same exception, 
and we got an average \emph{pyramid score} of 0.88. The finding indicates that the recommended queries are promising and comparable to the queries of expert users.
We also conducted experiments using 75 programming errors and exceptions, and compared our results against existing approaches \cite{context, seahawk}, and three traditional web search engines (\eg\ Google, Bing, Yahoo) and StackOverflow \cite{surf}. The proposed tool outperforms the existing approaches both in \emph{precision} and \emph{recall}. Among the search engines, Google performs the best. While our tool provides slightly less precise results than Google, it performs significantly better than Google in terms of \emph{recall}. Detailed results can be found elsewhere \cite{surf, scp}.



\section{Conclusion \& Future Works}\label{sec:conclusion}
To summarize, we propose a context-aware meta search solution, \emph{SurfClipse}, to the programming errors and exceptions encountered by software developers. 
The tool works both as a search query recommender and a meta search engine, and helps the developers in solving their programming problems especially associated with programming errors and exceptions.
In future, we plan to conduct a more exhausted user study with prospective participants.
We also plan for the recommendation of more sophisticated items such as relevant sections from a selected web page so that developers can easily locate the solutions and can solve the problems with reduced efforts.

\bibliographystyle{plainnat}
\scriptsize
\bibliography{sigproc}  
%
%

\end{document}